\documentclass{article}

\usepackage{PRIMEarxiv}

\usepackage[utf8]{inputenc} 
\usepackage[T1]{fontenc}    
\usepackage{url}   
\usepackage{bm}
\usepackage{sectsty}
\usepackage{nccmath}
\usepackage{amsmath}
\usepackage[numbers]{natbib}
\usepackage{dcolumn}
\usepackage{mathtools}
\usepackage{float}
\usepackage{fancyhdr}
\usepackage{fnpos}
\usepackage[english]{babel}
\usepackage{lastpage}
\usepackage{booktabs}       
\usepackage{amsfonts}
\usepackage{mathptmx}
\usepackage{mathtools}
\usepackage{amsfonts}
\makeatletter
\def\blfootnote{\xdef\@thefnmark{}\@footnotetext}
\makeatother
\usepackage{nicefrac}       
\usepackage{microtype}      
\usepackage{lipsum}
\usepackage{fancyhdr}       
\usepackage{graphicx}       
\graphicspath{{media/}}     

\title{First-principle validation of Fourier's law in \ensuremath{d=1,2,3}  classical systems

}

\author{
   \noindent\large{Constantino Tsallis\textit{$^{a}$}, Henrique Santos Lima\textit{$^b$}, Ugur Tirnakli\textit{$^{c}$} and Deniz Eroglu\textit{$^{d}$}}
}

\begin{document}
\maketitle

\begin{flushright}\vspace{-1.0cm}\textit{\textbf{``Yo soy yo y mi circunstancia''}} (Ortega y Gasset)\end{flushright} \vspace{0.5cm}

\begin{abstract}
We numerically study the thermal transport in the classical inertial nearest-neighbor XY  ferromagnet in $d=1,2,3$, the total number of sites being given by $N=L^d$, where $L$ is the linear size of the system.  For the thermal conductance $\sigma$, we obtain  $\sigma(T,L)\, L^{\delta(d)} = A(d)\, e_{q(d)}^{- B(d)\,[L^{\gamma(d)}T]^{\eta(d)}}$ (with $e_q^z \equiv [1+(1-q)z]^{1/(1-q)};\,e_1^z=e^z;\,A(d)>0;\,B(d)>0;\,q(d)>1;\,\eta(d)>2;\,\delta \ge 0; \,\gamma(d)>0)$, for all values of $L^{\gamma(d)}T$ for $d=1,2,3$. In the $L\to\infty$ limit, we have $\sigma \propto 1/L^{\rho_\sigma(d)}$ with  $\rho_\sigma(d)= \delta(d)+ \gamma(d) \eta(d)/[q(d)-1]$. The material conductivity is given by $\kappa=\sigma L^d \propto 1/L^{\rho_\kappa(d)}$ ($L\to\infty$) with $\rho_\kappa(d)=\rho_\sigma(d)-d$. Our numerical results are consistent with 'conspiratory' $d$-dependences of $(q,\eta,\delta,\gamma)$, which comply with normal thermal conductivity (Fourier law) for all dimensions. 
\end{abstract}

\blfootnote{\textit{$^{a}$~Centro Brasileiro de Pesquisas Fisicas and National Institute of Science and Technology of Complex Systems, Rua Xavier Sigaud 150, Rio de Janeiro-RJ 22290-180, Brazil \\
Santa Fe Institute, 1399 Hyde Park Road, Santa Fe, 
 New Mexico 87501, USA \\
Complexity Science Hub Vienna, Josefst\"adter Strasse 
 39, 1080 Vienna, Austria \\
E-mail: tsallis@cbpf.br}}

\blfootnote{\textit{$^{b}$~Centro Brasileiro de Pesquisas Fisicas, Rua Xavier Sigaud 150, Rio de Janeiro-RJ 22290-180, Brazil.\\ 
E-mail: hslima94@cbpf.br }}
\blfootnote{\textit{$^{c}$~Department of Physics, Faculty of Arts and Sciences, Izmir University of Economics, 35330, Izmir, Turkey. \\ E-mail: ugur.tirnakli@ege.edu.tr }}
\blfootnote{\textit{$^{d}$~Faculty of Engineering and Natural Sciences, Kadir Has University, 34083, Istanbul, Turkey \\ E-mail: deniz.eroglu@khas.edu.tr }}

\section{Introduction}
Fourier's law \cite{Fourier1822}  describes the heat diffusion rate through a macroscopic material in the direction of the flow. It accurately illustrates linear thermal transport at the macro-scale, and relies on the fundamental assumption of local thermal equilibrium. At the nano-scale, the experimental and theoretical studies on heat conduction revealed the emergence of new behaviors due to interactions and geometry, giving rise to novel material features representing unforeseen technical possibilities in non-equilibrium phenomena. The ability to regulate the behavior of heat flux in such cases is crucial for predicting, thus, controlling these systems' behavior to acquire the desired functionality and design new technologies. Therefore, a strong understanding of the interaction and geometrical effects on fundamental and predictive thermal transport properties is important for future technologies.

The reason for the unexpected heat flow behavior regards the interactions is known as the energy-carrying mechanisms of the systems. For instance, the Fourier's law  is invalid where the energy dissipation is related to the collisions such as fluidized granular media in classical systems \cite{soto1999} or where phonons are dominantly involved as heat carriers, such as dielectric and semiconductor materials, due to the definition of thermal equilibrium is ill-posed for phonon–phonon scattering in quantum systems~\cite{chen1996,MingoBroido2005,Jiangetal2009,siemens2010,Yangetal2015}. The effect of system size on low-dimensional systems is also theoretically studied \cite{Peter1998,Leprietal2003,LiCasatietal2004,Yangetal2006,Dhar2008,Leetal2017}, and the thermal conductivity of low-dimensional momentum conserving systems showed a system-size-dependent abnormality in the thermodynamic limit \cite{mendl2013, vanbeijeren2012, chen2016, Luo2020}, meanwhile, finite-size \cite{MejiaMonasterioProsenCasati2005,mejia2019heat} and momentum preserving systems have no abnormality \cite{savin2014, zhong2013}. However, real materials have a finite system size and three-dimensional geometry. Therefore, it is still an essential and open question to numerically reveal the topology effect where the system size is finite, and the interactions do not break Fourier's law. 

A classical and pragmatic XY model (or planar-rotator model), in $d$-dimensional hypercubic lattices, is selected to evaluate the validity of Fourier's law. The XY model was studied in the 1-dimensional case for low temperatures, and the change in conductivity concerning temperature was satisfactorily fitted with $q$-Gaussian distributions \cite{LiLiTirnakliLiTsallis2017}. In this article, we studied the XY model for $d=1,2$ and $3$ dimensional cases, which allowed us to evaluate the validity of the Fourier's law. Furthermore, we better characterized the conductivity change for a more extended range of temperatures, resulting in the $q$-stretched exponential instead of the $q$-Gaussian distribution. 
\section{Model}
 
\begin{figure}
\centering
\hspace{1.5cm}
  \includegraphics[width=10cm]{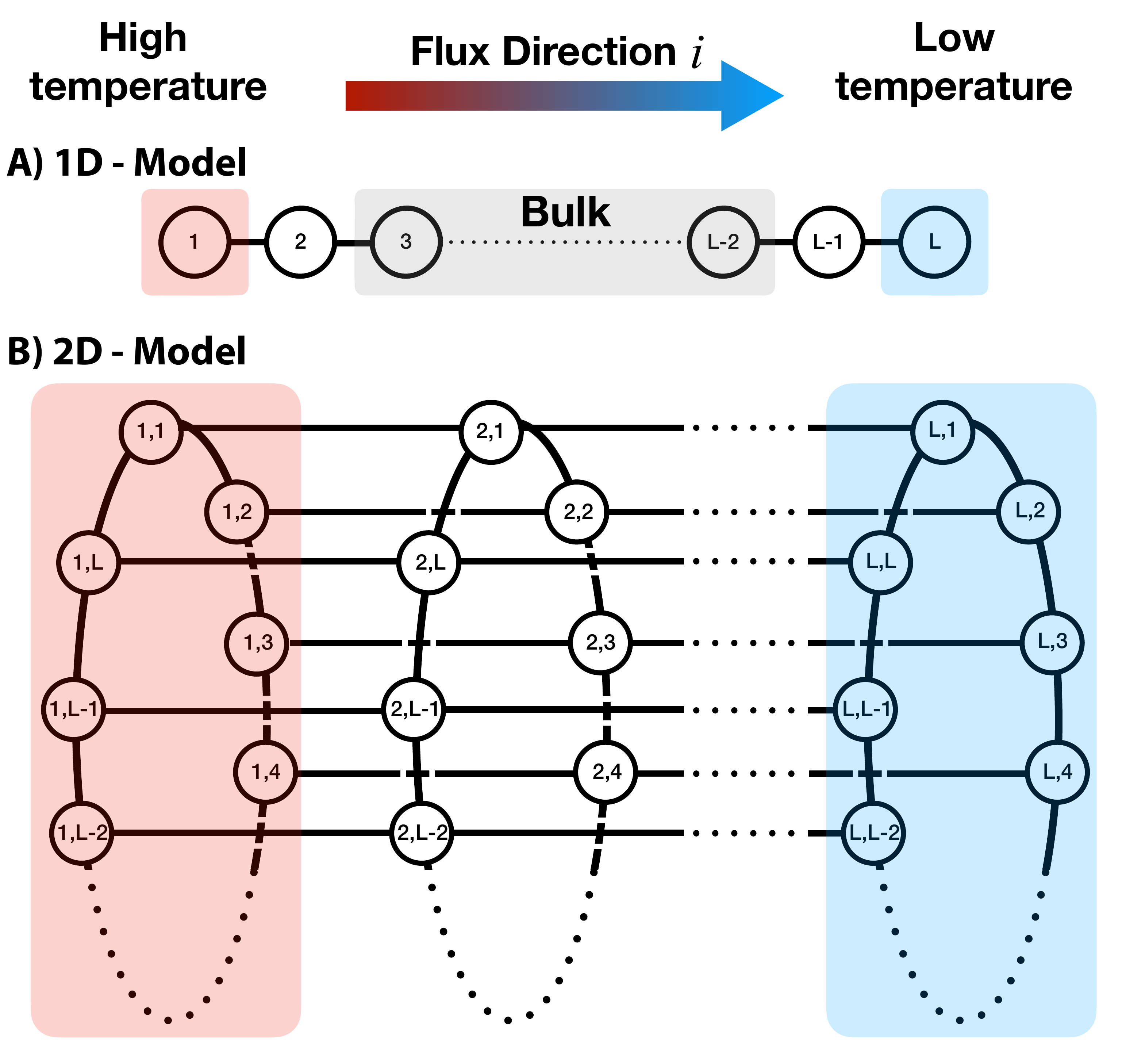}
  \caption{The lattice structure of the present A) $d=1$ model ($L$ sites) and B) $d=2$ model ($L^2$ sites). Red shaded areas represent hot heat bath, $T_h$, and blue areas are cold heat bath, $T_l$. The heat flux direction is from the hot heat bath to the cold one. To sensitively compute the heat flux and conductance, the bulk selected from the 3rd component to $L-2$ one in the flux direction to avoid direct random noise from the heat baths. The bulk is illustrated for 1D-model in A), which is straightforwardly generalized for dimensions $d$=2 and 3.} 
  \label{1dModel}
\end{figure}

The Hamiltonian of the $d$-dimensional inertial ferromagnetic XY model is given by 
\begin{ceqn}
\begin{eqnarray}
{\cal H}= \frac{1}{2}\sum_{\ell=1}^{L^d} p_{\ell}^2 +\frac{1}{2}\sum_{\langle \ell,\ell^\prime \rangle} [1- \cos(\theta_{\ell }-\theta_{\ell^\prime})] \,,
\label{XYHamiltonian}
\end{eqnarray}
\end{ceqn}
where $\langle \ell,\ell^\prime \rangle$ denotes nearest-neighboring rotors in the $d$-dimensional lattice \cite{LiLiTirnakliLiTsallis2017,LiLiLi2015,OlivaresAnteneodo2016}.  Because we assume that the particles have the same mass and the same moment of inertia, we have considered unit momenta of inertia and unit first-neighbor coupling constant without loss of generality, and $(p_{\ell},\theta_{\ell})$ are conjugate canonical pairs. We use  periodic boundary conditions along $(d-1)$ directions, and leaving open for $1$-dimensional ends. One of the ends being at a low temperature heat bath $T_l$ and the other one at high temperature $T_h$ (see Fig.~\ref{1dModel} for the illustration for $d=1$ and $2$).

 The equation of motion for the one-dimensional model is given as,

\begin{ceqn}
\begin{align}
\begin{split}
\dot \theta_i&=p_i\,\,\,\text{($i=1, \dots, L$)}\\
    \dot p_1&=-\gamma_h p_1+F_1+\sqrt{2\gamma_h T_h}\eta_h(t)\\
    \dot p_i &=F_i \,\,\,\text{($i=2,\dots, L-1$)}\\
    \dot p_L&=-\gamma_l p_L+F_L+\sqrt{2\gamma_l T_l}\eta_l(t) \,,\\
         \end{split}
   \end{align}
   \end{ceqn}
the force components being given by
\begin{ceqn}
\begin{align}
\begin{split}
 F_1&= -\sin(\theta_1-\theta_{2})-\sin(\theta_1)\\
   F_i&= -\sin(\theta_i-\theta_{i+1})-\sin(\theta_i-\theta_{i-1})\\
    F_L&= -\sin(\theta_L)-\sin(\theta_L-\theta_{L-1}),
    \end{split}
\end{align}
\end{ceqn}
where $i=2,\dots, L-1$, the friction coefficients are chosen $\gamma_l=\gamma_h=1$ (for numerical convenience), and $\eta_l$ and $\eta_h$ represents the Gaussian white noise with zero mean value and unit variance. Note that, in a relativistic context, these equations must be modified.

\subsection{Equations of motion for $d>1$ Lattices}\vspace{0.5cm}

\subsubsection{2-Dimensional Lattice}

The equations of motion for $d=2$ are written as follows
\begin{ceqn}
\begin{align}
\begin{split}
\dot \theta_{i,j}&=p_{i,j}\,\,\,\text{$((i,j)=1, \dots, L)$}\\
    \dot p_{1,j}&=-\gamma_h p_{1,j}+F_{1,j}+\sqrt{2\gamma_h T_h}\eta_{j,h}(t)\\
    \dot p_{i,j} &=F_{i,j} \,\,\,\text{($i=2,\dots, L-1$)}\\
    \dot p_{L,j}&=-\gamma_l p_{L,j}+F_{L,j}+\sqrt{2\gamma_l T_l}\eta_{j,l}(t)\,,\\
    \end{split}
\end{align}
\end{ceqn}

the force components being given by
\begin{ceqn}
\begin{align}
\begin{split}
F_{1,j}&= -\sin(\theta_{1,j}-\theta_{2,j})-\sin(\theta_{1,j})\\        
  & -\sin(\theta_{1,j}-\theta_{1,j+1})-\sin(\theta_{1,j}-\theta_{1,j-1})\\
  F_{i,j}&= -\sin(\theta_{i,j}-\theta_{i+1,j})-\sin(\theta_{i,j}-\theta_{i-1,j})\\        
  & -\sin(\theta_{i,j}-\theta_{i,j+1})-\sin(\theta_{i,j}-\theta_{i,j-1})\\
  F_{L,j}&= -\sin(\theta_{L,j})-\sin(\theta_{L,j}-\theta_{L-1,j})\\        
  & -\sin(\theta_{L,j}-\theta_{L,j+1})-\sin(\theta_{L,j}-\theta_{L,j-1})\\
    \end{split}
\end{align}
\end{ceqn}
where $\theta_{i,1}=\theta_{i,L+1}$ and $\theta_{i,0}=\theta_{i,L}$. The friction coefficients $\gamma_l$ and $\gamma_h$ have been  chosen $\gamma_l=\gamma_h=1$, and all components of the vectors  $\eta_{j,l}$ and $\eta_{j,h}$ are random Gaussian distributions with zero mean value and unit variance.

 \subsubsection{3-Dimensional Lattice}
For $d=3$, we have  similarly :
 \begin{ceqn}
\begin{align}
\begin{split}
\dot \theta_{i,j,k}&=p_{i,j,k}\,\,\,\text{$((i,j,k)=1, \dots, L)$}\\
    \dot p_{1,j,k}&=-\gamma_h p_{1,j,k}+F_{1,j,k}+\sqrt{2\gamma_h T_h}\eta_{j,k,h}(t)\\
    \dot p_{i,j,k} &=F_{i,j,k} \,\,\,\text{$(i=2,\dots, L-1)$}\\
    \dot p_{L,j,k}&=-\gamma_l p_{L,j,k}+F_{L,j,k}+\sqrt{2\gamma_l T_l}\eta_{j,k,l}(t)\,,\\
    \end{split}
\end{align}
\end{ceqn}
the force components being given by
\begin{ceqn}
\begin{align}
\begin{split}
F_{1,j,k}&= -\sin(\theta_{1,j,k}-\theta_{2,j,k})-\sin(\theta_{1,j,k})\\        & -\sin(\theta_{1,j,k}-\theta_{1,j+1,k})-\sin(\theta_{1,j,k}-\theta_{1,j-1,k})\\            &-\sin(\theta_{1,j,k}-\theta_{1,j,k+1})-\sin(\theta_{1,j,k}-\theta_{1,j,k-1})\\
F_{i,j,k}&= -\sin(\theta_{i,j,k}-\theta_{i+1,j,k})-\sin(\theta_{i,j,k}-\theta_{i-1,j,k})\\        & -\sin(\theta_{i,j,k}-\theta_{i,j+1,k})-\sin(\theta_{i,j,k}-\theta_{i,j-1,k})\\
            &-\sin(\theta_{i,j,k}-\theta_{i,j,k+1})-\sin(\theta_{i,j,k}-\theta_{i,j,k-1})\\
F_{L,j,k}&= -\sin(\theta_{L,j,k})-\sin(\theta_{L,j,k}-\theta_{L-1,j,k})\\        & -\sin(\theta_{L,j,k}-\theta_{L,j+1,k})-\sin(\theta_{L,j,k}-\theta_{L,j-1,k})\\
            &-\sin(\theta_{L,j,k}-\theta_{L,j,k+1})-\sin(\theta_{L,j,k}-\theta_{L,j,k-1})\\          
    \end{split}
\end{align}
\end{ceqn}
where $\theta_{i,1,k}=\theta_{i,L+1,k}, \theta_{i,0,k}=\theta_{i,L,k}, \theta_{i,j,1}=\theta_{i,j,L+1}$
        and $\theta_{i,j,0}=\theta_{i,j,L}$.
 The friction coefficients $\gamma_l$ and $\gamma_h$ have been chosen $\gamma_l=\gamma_h=1$, and all components of the matrices  $\eta_{j,k,l}$ and $\eta_{j,k,h}$ are random Gaussian distributions with zero mean value and unit variance.
\onecolumn
\subsection{Arbitrary Interaction Topology}
The equations of the motion of $N$ interacting rotors for any interaction topology can also be written in a compact form as follows:
\begin{ceqn}
\begin{align}
\begin{split}
\dot{\theta}_{i} &= p_{i} \\
\dot{p}_{i} &= 
\begin{dcases} 
-\gamma_h p_{i} - \sin(\theta_{i}) -\sum_{j=1}^{N} A_{ij}\sin(\theta_{i} - \theta_{j}) + \sqrt{2\gamma_hT_h}\eta_{i}(t) 
& : i \in R_h\\ 
-\sum_{j=1}^{N} A_{ij}\sin(\theta_{i} - \theta_{j})
& : i \in R_b\\
 -\gamma_l p_{i} -\sin(\theta_{i}) - \sum_{j=1}^{N} A_{ij}\sin(\theta_{i} - \theta_{j}) + \sqrt{2\gamma_lT_l}\mu_{i}(t) 
 & : i \in R_l
\end{dcases}
\end{split}
\end{align}
\end{ceqn}

where $\mathbf{A}=[A_{ij}]$ is the topological interaction matrix, and $R_h, R_b$ and $R_l$ are the sets of rotors in the high-temperature heat bath $T_h$, the bulk and the low-temperature heat bath $T_l$, respectively. The associated lattice topology matrices, $\mathbf{A}$, we use in the current work for $d=1,2,3$ can be rewritten as follows:\vspace{0.5cm}
\subsubsection
{1d-Model: Chain Topology}
The connectivity matrix, $\bm A_{\text{chain}}$is $L \times L$ matrix representing a 1-dimensional chain system, defined as follows:
\begin{ceqn}
\begin{eqnarray}
\bm A_{\text{chain}} = 
\begin{bmatrix}
0 & 1 &  &  & &  \\
1 & 0 & 1 &  & {\text{\huge0}} &\\
 & 1 & 0 & \ddots & & \\
 & & \ddots & \ddots & 1 & \\
& \makebox(0,0){\text{\huge0}} &  &1&0&1\\
&&&&1&0
\end{bmatrix}
\end{eqnarray}
\end{ceqn}

\subsubsection{2d-Model: Cylinder Topology}
The connectivity matrix, $\bm A_{\text{cylinder}}$, is $L^2 \times L^2$ matrix representing a 2-dimensional lattice system with periodic boundary conditions through one axis (cylinder shape), which is defined as follows:
\begin{ceqn}
\begin{eqnarray}
\bm A_{\text{cylinder}} = 
\begin{bmatrix}
\bm A_{\text{ring}} & \bm I &  &  & & \\
\bm I & \bm A_{\text{ring}} & \bm I &  & {\text{\huge0}} &\\
 & \bm I & \bm A_{\text{ring}} & \ddots & & \\
 & & \ddots & \ddots & \bm I & \\
& \makebox(0,0){\text{\huge0}} &  &\bm I&\bm A_{\text{ring}}&\bm I\\
&&&&\bm I& \bm A_{\text{ring}} 
\end{bmatrix}
\end{eqnarray}
\end{ceqn}
where $\bf I$ is $L \times L$ identity matrix and $\bm A_{\text{ring}}$ is $L \times L$ matrix as follows:
\begin{ceqn}
\begin{eqnarray}
\bm A_{\text{ring}} = 
\begin{bmatrix}
0 & 1 &  &  & & 1 \\
1 & 0 & 1 &  & {\text{\huge0}} &\\
 & 1 & 0 & \ddots & & \\
 & & \ddots & \ddots & 1 & \\
& \makebox(0,0){\text{\huge0}} &  &1&0&1\\
1 &&&&1&0
\end{bmatrix}
\end{eqnarray}
\end{ceqn}

\subsubsection{3d-Model: Coupled-Tori Topology}
The connectivity matrix, $\bm A_{\text{coupled-tori}}$, is $L^3 \times L^3$ matrix representing 3-dimensional coupled tori system, which is defined as follows:
\begin{ceqn}
\begin{eqnarray}
\bm A_{\text{coupled-tori}} = 
\begin{bmatrix}
\bm A_{\text{torus}} & \bm I_2 &  &  & & \\
\bm I_2 &  {\bm A_{\text{torus}}} & \bm I_2 &  & {\text{\huge0}} &\\
 & \bm I_2 & {\bm A_{\text{torus}}} & \ddots & & \\
 & & \ddots & \ddots & \bm I_2 & \\
& \makebox(0,0){\text{\huge0}} &  &\bm I_2& {\bm A_{\text{torus}}}&\bm I_2\\
&&&&\bm I_2&\bm A_{\text{torus}} 
\end{bmatrix}
\end{eqnarray}
\end{ceqn}
where $\bm I_2$ is $L^2 \times L^2$ identity matrix  and $\bm  A_{\text{torus}}$ is $L^2 \times L^2$ matrix as follows:
\begin{ceqn}
\begin{eqnarray}
\bm A_{\text{torus}} = 
\begin{bmatrix}
\bm A_{\text{ring}} & \bm I &  &  & & \bm I\\
\bm I & \bm A_{\text{ring}} & \bm I &  & {\text{\huge0}} &\\
 & \bm I & \bm A_{\text{ring}} & \ddots & & \\
 & & \ddots & \ddots & \bm I & \\
& \makebox(0,0){\text{\huge0}} &  &\bm I&\bm A_{\text{ring}}&\bm I\\
\bm I&&&&\bm I& \bm A_{\text{ring}}
\end{bmatrix}.
\end{eqnarray}
\end{ceqn}
\twocolumn
\section{Methods}
The dynamical evolution was conducted using  the Velocity-Verlet algorithm \cite{Verlet1967,PaterliniFerguson1998} with  step size $dt=0.01$; after discarding a transient time, the average of the heat flux is computed for $4\times 10^8$ time steps and $80$ randomly initialized realizations. The transient time is carefully selected for different system sizes by considering the development of the conductivity curve for varying temperature values. The system is assumed to be stationary when the conductivity curve reaches a steady state. For simplicity, we set  $T_{h}=T(1+ \Delta)$ and $T_{l}=T(1- \Delta)$ with $\Delta=0.125$, where $T$ is the average temperature . The macroscopic conductivity $\kappa$ is given by

\begin{ceqn}
\begin{equation}
    \kappa= \frac{J}{(T_h-T_l)/L} =\frac{J}{2\Delta T/L}
\end{equation}
\end{ceqn}
where $J = \langle J_l\rangle_{bulk}$ is the time and space average of heat flux along the bulk of the lattice in the stationary state, which connects the microscopic level (the equations of motion) with the macroscopic one (the average of the heat flux and thermal conductivity) via the continuity equation. The bulk is defined as the entire system excluding the sides in high and low temperature heat baths and their first neighbors to avoid the direct effect of stochastic dynamics on the flux calculation (see Fig.~\ref{1dModel}).  Therefore, the possible minimum system length for any lattice topology $L^d$ is $L=5$ to compute the flow as desired. Furthermore, to reduce the direct effect of noise on the flux, one can ignore more than two nearest neighbors to the heat baths from the calculation for large systems.  The time derivative of the Hamiltonian Eq.~\ref{XYHamiltonian} can be written as 
\begin{ceqn}
\begin{equation}
    \frac{d\cal{H}}{dt}= -\frac{1}{2} \sum_{\ell=1}^{L^d}(J_{\ell}-J_{\ell ^\prime})
    \end{equation}
    \end{ceqn}
  
where $J_\ell = (p_\ell + p_{\ell^\prime})\sin(\theta_{\ell} - \theta_{\ell^\prime})$ is the Lagrangian flux \cite{mejia2019heat}, $\ell \in \{1, \cdots, L^d\}$ is a unique label for each site and $\ell^\prime$ is the nearest-neighbor of site-$\ell$ towards to hot reservoir. Therefore, $J_{\ell}$ is defined as the energy transfer per unit time, per transverse $(d-1)$-dimensional ``area" $L^{d-1}$. Note that the calculation of $J_{\ell}$ is independent of the lattice dimension $d$ since the flow direction is always in one direction from the high temperature end to the cold one. The statement for the flux direction is straightforward for $d=1$; however, the model for  $d>1$ has periodic boundary conditions for interacting sides on $(d-1)$ dimensions. Then the flux is defined only through the axis where the boundaries are ended with the heat baths in any lattice dimension $d \in \mathbb{Z}^+$. The macroscopic  conductivity $\kappa$ only depends on the specific material and its temperature. This is essentially the content of Fourier's 1822 law, where only the macroscopic phenomenon was considered~\cite{Fourier1822}.

The (dimensionless) conductivity $\kappa$ and the  (dimensionless) ``conductance" $\sigma$ are, by definition, related through
\begin{ceqn}
\begin{equation}
\kappa \equiv \sigma L^d \,.
\end{equation}
\end{ceqn}
As we shall later on verify, this specific definition of $\sigma$ \cite{LiLiTirnakliLiTsallis2017} does not depend, for $d=1$, on $L$ in the $T\to 0$ limit (see Fig. 2).

The asymptotic power-law relation between $T$ and $\sigma$ (or $\kappa$) was numerically explored for the one-dimensional first-neighbor planar-rotator model \cite{LiLiLi2015}. Furthermore,  a collapse of the power-law distributions was discovered through the following $q$-Gaussian~\cite{LiLiTirnakliLiTsallis2017}
\begin{ceqn}
\begin{equation}
\sigma(T,L) = \sigma(0,L) \, e_q^{-B_q (L^{1/3}T)^2}\,, 
\label{oldqgaussian}
\end{equation}
\end{ceqn}
where, for $d=1$, $\sigma(0,L)$ is independent from $L$, and $(q,B_q)\simeq (1.55,0.40)$, the $q$-exponential function being defined as $e_q^z \equiv [1+ (1-q)z]^{1/(1-q)}$ ($e_1^z=e^z$).The $q$-Gaussian form (\ref{oldqgaussian})
was proposed in \cite{LiLiTirnakliLiTsallis2017} because, under appropriate simple constraints, it extremizes the nonadditive entropy
\begin{ceqn}
\begin{eqnarray}
S_q &\equiv& k\,\frac{1-\sum_ip_i^q}{q-1}=k\,\sum_ip_i \ln_q\frac{1}{p_i} \nonumber \\
&=&-k\,\sum_i p_i^q \ln_q  p_i= -k\,\sum_i p_i \ln_{2-q}p_i 
\end{eqnarray}
\end{ceqn}
where $k$ is a positive constant such that for $q=1$, $k=k_B$ ($k_B$ is the Boltzmann constant), and  $\ln_q z \equiv \frac{z^{1-q}-1}{1-q}\;(\ln_1 z =\ln z)$ \cite{Tsallis1988,Tsallis2009,Tsallis2022}. We straightforwardly verify that $S_1=S_{BG}\equiv -k \, \sum_i p_i \ln p_i$, where BG stands for Boltzmann-Gibbs. We also verify that, for two statistically  independent systems $X$ and $Y$ (i.e., $p_{ij}^{X+Y}=p_i^X p_j^Y $),
\begin{ceqn}
\begin{equation}
\frac{S_q(X+Y)}{k}=\frac{S_q(X)}{k}+\frac{S_q(Y)}{k}+(1-q)\frac{S_q(X)}{k}\frac{S_q(Y)}{k} \,.    
\end{equation}
\end{ceqn}
This property exhibits the nonadditivity of the entropic functional $S_q$ for $q \ne 1$. For $q=1$ we recover the well known BG additivity $S_{BG}(X+Y)=S_{BG}(X)+S_{BG}(Y)  $, which follows Penrose's definition of entropic additivity \cite{Penrose1970}.
\section{Results}
 We revisit here the $d=1$ results of \cite{LiLiTirnakliLiTsallis2017} by exploring higher values of $T$. It turns out that, while the $q$-Gaussian Ansatz was good enough for the conductivity $\sigma$ at the relatively low temperatures considered in \cite{LiLiTirnakliLiTsallis2017}, the present numerics at a wider range of $T$ require a more general Ansatz, namely the stretched $q$-exponential
 \begin{ceqn}
\begin{equation}
y(x)=e_q^{-B|x|^\eta}\,
\label{moregeneral}
\end{equation}
\end{ceqn}
with $q\ge 1$, $\eta > 0$ and $B>0$. The $q$-Gaussian form Eq.~(\ref{oldqgaussian}) is recovered as the $\eta=2$ particular limit of this more general form. The form Eq.~(\ref{moregeneral}) introduces one more parameter, namely $\eta$, which fits our numerical data very satisfactorily. Note that we used the standard least squares method to find the best-fitting curve for our numerical data. By so doing, we follow the successful Ansatz proposed in \cite{Grenoble} for neutron experiments with standard spin glasses. This is specifically shown in what follows here below. 

\begin{figure*}
\centering
\includegraphics[width=\textwidth]{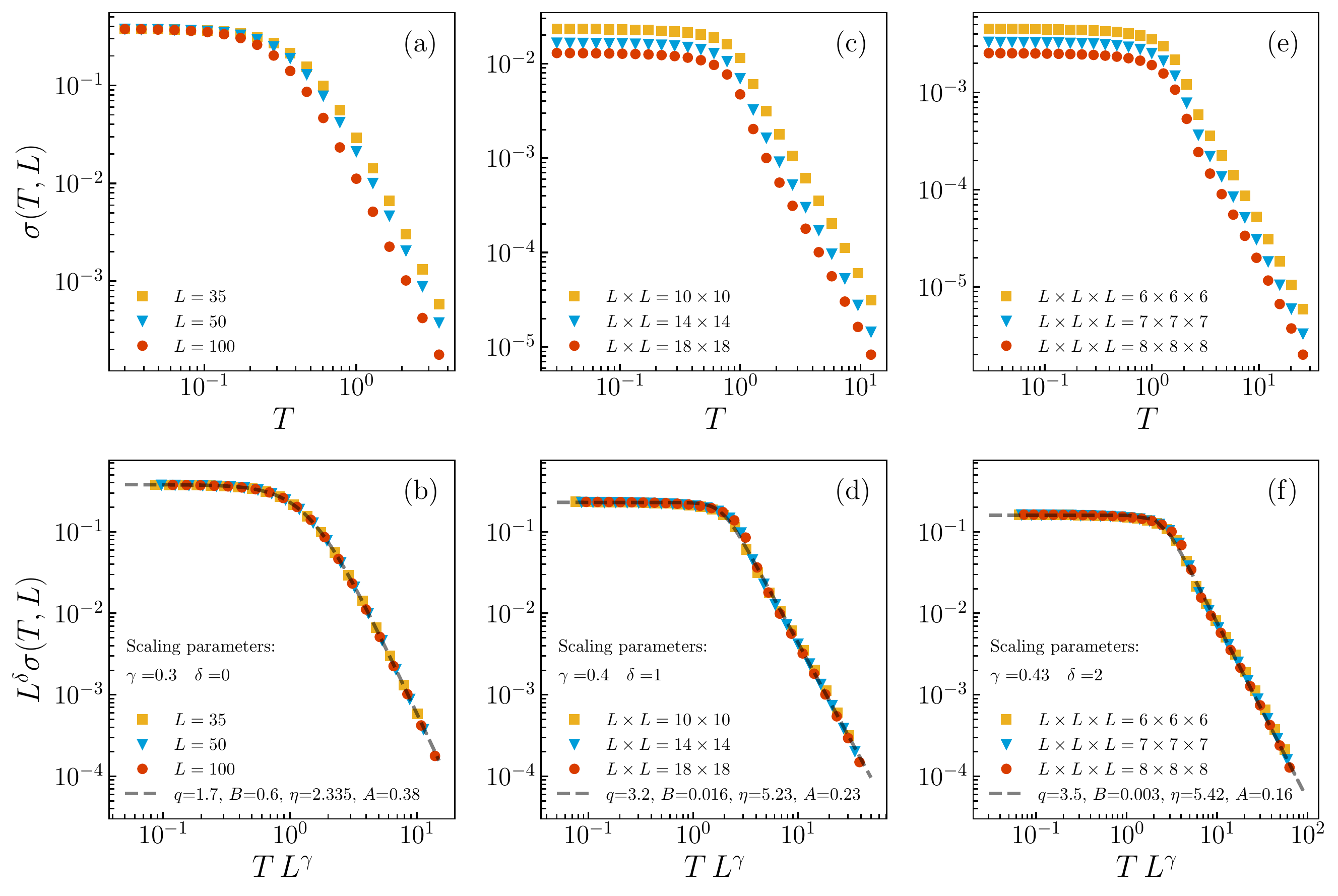}
  \caption{Thermal conductance as a function of temperature for $d$-dimensional lattice structures ($d=1,2$ and $3$). {\it Top:} Conductance $\sigma$ plotted for (a) dimension $d=1$ for sizes $L=35, 50$ and $100$, (c) $d=2$ with $L\times L =10\times 10, 14\times 14$ and $18\times18$ and (e) $d=3$ with $L\times L \times L = 6 \times 6 \times 6, 7 \times 7 \times7$ and $8 \times 8 \times8$. {\it Bottom:} Collapse of $ \sigma$ values for all available system sizes in dimensions (b) $d=1$, (d) $d=2$ and (f) $d=3$ using the relations for temperature $T \to TL^\gamma$ and $\sigma \to \sigma L^\delta$, scaling parameters, $\delta$ and $\gamma$, are given on the associated sub-figures. Collapsed $\sigma$ values are accurately fitted with $\sigma(T,L) = A(1-(1-q)B(TL^\gamma)^\eta)^{1/(1-q)}$ using the optimal parameters in the legend for the fitting curves (dashed gray lines).
  The number of time steps used for all $d$ case is $4\times 10^{8}$ and an average is taken over 80 experiments. The number of transients thrown away for the system to attain the stationary state is at least $2.6\times 10^{11}$ for $d=1$, $8.0\times 10^{10}$ for $d=2$ and $5.6\times 10^{10}$ for $d=3$.}
  \label{1d}
\end{figure*}

\begin{figure}
 \centering
 \includegraphics[width=\columnwidth]{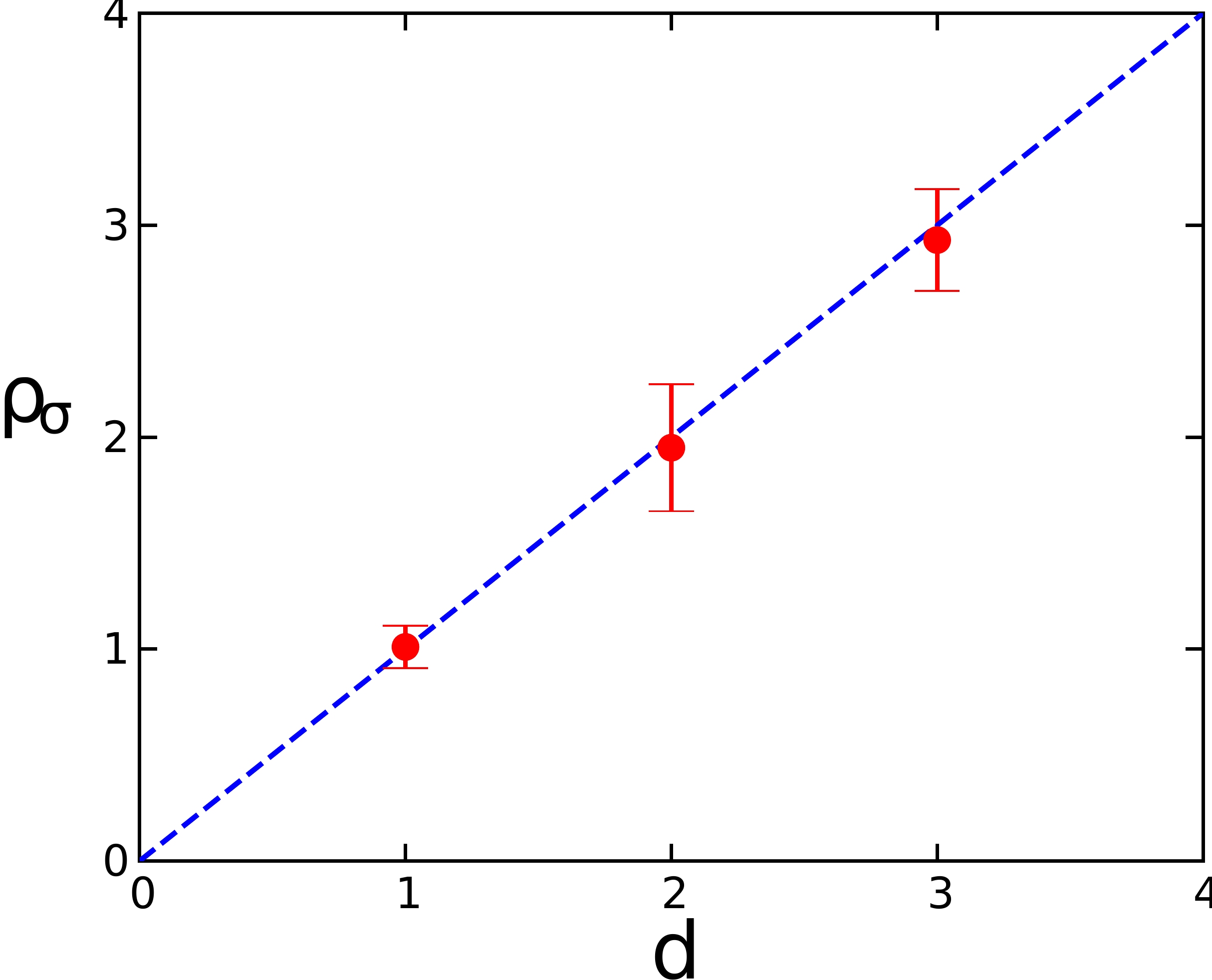}
 \caption{$\sigma \propto 1/L^{\rho_\sigma(d)}\;(L\to\infty)$ and $\kappa=\sigma L ^d \propto L ^{d-\rho_\sigma(d)}$. The dots correspond to the present numerical results. The dashed line indicates the validity of Fourier's law, i.e., $\lim_{L\to\infty} \kappa(T,L)$ is a finite $T$-dependent quantity. These results strongly suggest that $\rho_\sigma=d$, hence $\rho_\kappa=0$,  for all values of $d$, possibly including noninteger values as well.
 } 
 \label{fgr:example2col}
\end{figure}

All our results for $d=1,2$ and $3$ collapse in the following universal form:
\begin{ceqn}
\begin{equation}
\label{stretch}
\sigma(T,L)\,L^{\delta(d)}=
A(d)\,e_{q(d)}^{-B(d)[T\,L^{\gamma(d)}]^{\eta(d)}} \,,
\end{equation}
\end{ceqn}
where $(A,B,q,\eta,\gamma, \delta)$ are fitting parameters (Fig.~\ref{1d}). Let us emphasize here that Fourier's law corresponds to the $L \to \infty$ limit of this equation, hence, both $\sigma$ and $\kappa$ decay with power laws, namely $\sigma \sim 1/L^{\rho_{\sigma}}$ and $\kappa \sim 1/L^{\rho_{\kappa}}$, where $\rho_{\sigma}\equiv\delta+\gamma\frac{\eta}{q-1}$ and $\rho_{\kappa}\equiv \rho_{\sigma}-d$ as exhibited in Fig.~\ref{fgr:example2col}. The validation of Fourier's law is confirmed if $\rho_{\kappa}=0$ or,  equivalently, $\rho_{\sigma}=d$, making the thermal conductivity independent of the lattice size.  

\section{Conclusions}
The specific statistical mechanics correctly describing a given physical many-body problem depends on various aspects, including the range of the interactions and the boundary conditions. The classical model that is being focused on here concerns short-range interactions. Therefore, it constitutes a typical situation that, at thermal equilibrium, is correctly approached within the celebrated the BG theory (i.e., $q=1$). This would naturally be the case if we had $T_h=T_l$. However, the present non-equilibrium phenomenon relevantly modifies the thermostatistics to be used. Indeed, the present numerical results strongly indicate $q \ne 1$, thus suggesting that, for its proper discussion, the use of nonadditive entropies becomes a must.

Consistently with the above, at  the $L\to\infty$ limit, a sort of remarkable numerical 'conspiracy' of the values of $(q,\eta,\gamma, \delta)$ which, in the realm of first-principle Newtonian calculations, eventually implies the validity (i.e., $\rho_\kappa =0$), at all dimensions $d$ ( see Fig.~\ref{fgr:example2col}), of the centennial Fourier macroscopic law for thermal transport. Interestingly enough, an important ingredient of this numerical 'conspiracy' is the fact that seemingly $\delta=d-1$ for all dimensions $d$. 

In view of the present results in the collapsed form, namely $L^{d-1}\sigma(T,L) \propto e_{q(d)}^{-B(d) [L^{\gamma(d)}T]^{\eta(d)}}$ $(B>0,q>1,\eta>0,\gamma>0)$, $q$-Gaussians are replaced by $q$-stretched-exponentials \cite{LiLiTirnakliLiTsallis2017,Grenoble} due to the fact that a wider range of values of $T$ is presently focused on. We also intend to have in the future a closer look onto the influence of long-range interactions \cite{OlivaresAnteneodo2016}, and check whether the $q$-stretched-exponential form is preserved. 

Last but not least, we can emphasize here that there is no such a thing as physical systems which are $q=1$ or $q\ne 1$ ones, or even something else.  We should always bear in mind that the statistical mechanics which satisfactorily describes a given system depends not only on the nature itself of the system but also on its circumstances. More precisely, the time scale which is focused on, the size-scale which is appropriate, the precision degree which has been adopted, the class of initial conditions which is applied, and finally the boundary conditions under which the system is placed.  Suppose the system is in  thermal equilibrium (more specifically, as mentioned, with $T_h=T_l$, i.e., $d$-dimensionally periodic boundary conditions, instead of the $(d-1)$-dimensionally periodic ones that have been used here). In that case, mild spatial and time energy fluctuations  are compatible with ergodicity, and therefore the BG theory applies for the present short-range interacting classical system. The same system in a stationary-state which is permanently forced out of equilibrium, seemingly has space-time energy fluctuations whose nature is turbulent-like, therefore driving the system out of usual ergodicity and out of the BG theory, into $q$-statistics.

\section*{Acknowledgments}
We acknowledge fruitful remarks by G. Benedek, E.P. Borges and S. Miret Artes, as well as partial financial support from CNPq and Faperj (Brazilian agencies).
The numerical calculations reported in this paper were partially performed at TUBITAK ULAKBIM, High Performance and Grid Computing Center (TRUBA resources).
U.T. is a member of the Science Academy, Bilim Akademisi, Turkey. D.E. was supported by the BAGEP Award of the Science Academy, Turkey.

\section*{Author contributions}

All four authors are responsible for the concept, design, execution, and physical interpretation of the research. 

 \section*{Declaration of Competing interests}
The authors declare no competing interests.


\end{document}